\batchmode
\makeatletter
\def\input@path{{C:/Users/patilomkarsudhir/Desktop/}}
\makeatother
\RequirePackage{fixltx2e}
\documentclass[english]{IEEEtran}
\pdfoutput=1 
\usepackage[T1]{fontenc}
\usepackage[latin9]{inputenc}
\usepackage{geometry}
\usepackage{amsmath}
\usepackage{amsthm}
\usepackage{amssymb}
\usepackage{stackrel}

\makeatletter
\theoremstyle{definition}
\newtheorem{assumption}{Assumption}
\theoremstyle{plain}
\newtheorem{thm}{\protect\theoremname}
\theoremstyle{remark}
\newtheorem{rem}{\protect\remarkname}
\theoremstyle{plain}
\newtheorem{lem}{\protect\lemmaname}
\theoremstyle{plain}
\newtheorem{cor}{\protect\corollaryname}

\usepackage{cite}

\makeatother

\usepackage{babel}
\providecommand{\corollaryname}{Corollary}
\providecommand{\lemmaname}{Lemma}
\providecommand{\remarkname}{Remark}
\providecommand{\theoremname}{Theorem}

\begin{document}
\title{Adaptive Control of Time-Varying Parameter Systems with Asymptotic
Tracking}
\author{Omkar Sudhir Patil, Runhan Sun, Shubhendu Bhasin and Warren E. Dixon\thanks{Omkar Sudhir Patil, Runhan Sun, and Warren E. Dixon are with the Department
of Mechanical and Aerospace Engineering, University of Florida, Gainesville
FL 32611-6250 USA. Email: \{patilomkarsudhir,runhansun,wdixon\}@ufl.edu.}\thanks{Shubhendu Bhasin is with the Department of Electrical Engineering,
Indian Institute of Technology Delhi, New Delhi, India (e-mail: sbhasin@ee.iitd.ac.in). }\thanks{This research is supported in part by NSF award number 1762829, Office
of Naval Research Grant N00014-13-1-0151, AFOSR award number FA9550-18-1-0109
and FA9550-19-1-0169. Any opinions, findings and conclusions or recommendations
expressed in this material are those of the author(s) and do not necessarily
reflect the views of the sponsoring agency.}}
\maketitle
\begin{abstract}
A continuous adaptive control design is developed for nonlinear dynamical
systems with linearly parameterizable uncertainty involving time-varying
uncertain parameters. The key feature of this design is a robust integral
of the sign of the error (RISE)-like term in the adaptation law which
compensates for potentially destabilizing terms in the closed-loop
error system arising from the time-varying nature of uncertain parameters.
A Lyapunov-based stability analysis ensures asymptotic tracking, and
boundedness of the closed-loop signals.
\end{abstract}

\thispagestyle{empty}
\pagenumbering{gobble}

\section{Introduction}

Adaptive control of nonlinear dynamical systems with time-varying
uncertain parameters is an open and practically relevant problem.
It has been well established that traditional gradient-based update
laws can compensate for constant unknown parameters yielding asymptotic
convergence. Moreover, the development of robust modifications of
such adaptive update laws result in uniformly ultimately bounded (UUB)
results for slowly varying parametric uncertainty using a Lyapunov-based
analysis, under the assumption of bounded parameters and their time-derivatives
(cf. \cite{Ioannou1996}). 

More recent results focus on tracking and parameter estimation performance
improvement, though still limited to a UUB result, using various adaptive
control approaches for systems with unknown time-varying parameters.
One such approach involves a fast adaptation law \cite{gaudio2019parameter},
where a matrix of time-varying learning rates is utilized to improve
the tracking and estimation performance under a finite excitation
condition. Another approach uses a set-theoretic control architecture
\cite{Arabi.Gruenwald.ea2018,Arabi.Yucelen2019,Arabi.Yucelen.ea2019}
to reject the effects of parameter variation, while restricting the
system error within a prescribed performance bound. While the aforementioned
approaches can potentially yield improved transient response, the
results still yield UUB error systems. 

Motivation exists to obtain asymptotic convergence of the tracking
error to zero, despite the time-varying nature of the uncertain parameters.
Robust adaptive control approaches such as \cite{Qu2002} yield asymptotic
adaptive tracking for systems with time-varying uncertain parameters;
however, such approaches exploit high-gain feedback based on worst-case
uncertainty, rather than an adaptive control approach that scales
to compensate for the uncertainty without using worst-case gains.
In \cite{Xu2004}, the iterative learning control result in \cite{Qu2002}
is extended to yield asymptotic tracking for systems with periodic
time-varying parameters with known periodicity. 

To the best of our knowledge, an asymptotic tracking result has not
been achieved for a generalized class of nonlinear systems with unknown
time-varying parameters, where the parameters are not necessarily
periodic. Asymptotic tracking is difficult to achieve for the time-varying
parameter case because the time-derivative of the parameter acts like
an unknown exogenous disturbance in the parameter estimation dynamics,
which is difficult to cancel with an adaptive update law in a Lyapunov-based
stability analysis.

To illustrate this problem, consider the scalar dynamical system\footnote{Note that the system $(\text{\ref{eq:illussystem}})$ is considered
only for illustrative purpose. This paper presents result for a general
system with a vector state and a linearly parameterizable uncertainty
with time-varying parameters.} 

\begin{equation}
\dot{x}(t)=a(t)x(t)+u(t),\label{eq:illussystem}
\end{equation}
with the controller $u(t)=-kx(t)-\hat{a}(t)x(t)$, where $k$ is a
positive constant gain, $a(t)$ is the unknown time-varying parameter,
$\hat{a}(t)$ is the parameter estimate of $a(t)$ and the parameter
estimation error $\tilde{a}(t)$ is defined as $\tilde{a}(t)\triangleq a(t)-\hat{a}(t)$.
The traditional stability analysis approach for such problems is to
consider the Lyapunov function candidate $V(x(t),\tilde{a}(t))=\frac{1}{2}x^{2}(t)+\frac{1}{2\gamma}\tilde{a}^{2}(t)$,
where $\gamma$ is a positive constant gain. The given definitions
and controller yield the following time-derivative of the candidate
Lyapunov function: $\dot{V}(t)=-kx^{2}(t)+\tilde{a}(t)x^{2}(t)+\frac{\tilde{a}(t)}{\gamma}(\dot{a}(t)-\dot{\hat{a}}(t))$.
For the constant parameter case, i.e., $\dot{a}(t)=0$, the well-known
adaptive update law $\dot{\hat{a}}(t)=\gamma x^{2}(t)$ will cancel
the cross term $\tilde{a}(t)x^{2}(t)$ in $\dot{V}(t)$. However,
when the parameters are time-varying, it is unclear how to cancel
or dominate $\dot{a}(t)$ via an update law such that $\dot{V}(t)$
becomes at least negative semi-definite. It would be desirable to
have a sliding mode-like term based on $\tilde{a}(t)$ in the adaptation
law, but $\tilde{a}(t)$ is unknown. Another approach could be to
use a robust controller, e.g., $u(t)=-kx(t)-\bar{a}x(t)$, where $\bar{a}$
is a known constant upper bound on the norm of parameter $\left|a(t)\right|$,
or an adaptive robust controller which involves certainty equivalence
in terms of the unknown bound $\bar{a}$. Either of these approaches
would yield an asymptotic tracking result (cf., \cite{Qu2002}), but,
as stated earlier, these approaches are based on a high-gain worst
case scenario, rather than an adaptive control approach that scales
to compensate for the uncertainty without using worst-case gains.

A popular approach to design adaptive controllers for the time-varying
parameter case is to consider robust modification of the update law
and assume upper bounds of $\left|a(t)\right|$ and $\dot{a}(t)$
to obtain a UUB result. For instance, consider a standard gradient
update law with sigma-modification \cite{Ioannou.Kokotovic1983},
$\dot{\hat{a}}(t)=\gamma x^{2}(t)-\gamma\sigma\hat{a}(t)$, which
yields $\dot{V}(t)=-kx^{2}(t)-\sigma\tilde{a}^{2}(t)+\tilde{a}(t)(\frac{\dot{a}(t)}{\gamma}+\sigma a(t))$,
implying a UUB result when the parameter $a(t)$ and its time-derivative
$\dot{a}(t)$ are bounded. Moreover, the approaches developed in results
such as \cite{gaudio2019parameter} and \cite{Arabi.Yucelen2019}
can be used to improve the transient response of the UUB error system.

The major challenge to achieve asymptotic stability is the derivative
of the time-varying parameter term in the Lyapunov analysis, which
is addressed in this paper with a Lyapunov-based design approach,
that is inspired by the modular adaptive control approach in \cite{Patre2011}.
This approach includes higher order dynamics which appear after taking
the time-derivative of $\text{(\ref{eq:illussystem})}$. Since these
higher order dynamics contain the time-derivative of the parameter
estimate $\dot{\hat{a}}(t),$ it is possible to design $\dot{\hat{a}}(t)$
to facilitate the subsequent stability analysis. With this motivation,
a continuous adaptive control algorithm is developed for nonlinear
dynamical systems with linearly parameterized uncertainty involving
time-varying parameters, where a semi-global asymptotic tracking result
is achieved. A key feature of the proposed method is a robust integral
of the sign of the error (RISE)-like (see \cite{Xian2004,Makkar2007,Patre2008a,Patre2011})
update law, i.e., the update law contains a signum function of the
tracking error term multiplied by some desired regressor based terms.
The update law also involves a projection algorithm to ensure that
the parameter estimates stay within a bounded set. However, the projection
algorithm introduces a potentially destabilizing term in the time-derivative
of the Lyapunov function candidate, leading to an additional technical
obstacle to obtain asymptotic tracking. This challenge is resolved
by using an auxiliary term in the control input, which facilitates
stability by providing a stabilizing term and canceling the aforementioned
potentially destabilizing term in the time-derivative of the candidate
Lyapunov function. With the proposed method, the closed-loop system
dynamics have the same structure as previous RISE controllers \cite{Xian2004,Makkar2007,Patre2008a,Patre2011},
for which the stability analysis tools are well established, yielding
asymptotic convergence of the tracking error to zero, boundedness
of the parameter estimation error, and boundedness of the closed-loop
signals. 

\section{Dynamic Model}

Consider a control affine system with the nonlinear dynamics

\begin{equation}
\dot{x}(t)=h(x(t),t)+d(t)+u(t),\label{eq:dynamicsystem}
\end{equation}
where $x:[0,\infty)\to\mathbb{R}^{n}$ denotes the state, $h:\mathbb{R}^{n}\times[0,\infty)\to\mathbb{R}^{n}$
denotes a continuously differentiable function, $d:[0,\infty)\to\mathbb{R}^{n}$
represents an exogenous disturbance acting on the system, and $u:[0,\infty)\to\mathbb{R}^{n}$
represents the control input. The function $h(x(t),t)$ in $(\text{\ref{eq:dynamicsystem}})$
is assumed to be linearly parameterized as
\begin{equation}
h(x(t),t)\triangleq Y_{h}(x(t),t)\theta_{f}(t),\label{eq:fparam}
\end{equation}

\noindent where $Y_{h}:\mathbb{R}^{n}\times[0,\infty)\to\mathbb{R}^{n\times m}$
is a known regression matrix, and $\theta_{f}:[0,\infty)\to\mathbb{R}^{m}$
is a vector of time-varying unknown parameters.

The disturbance parameter vector $d(t)$ can be appended to the $\theta_{f}(t)$
vector, yielding an augmented parameter vector $\theta:[0,\infty)\to\mathbb{R}^{n+m}$
as 

\begin{equation}
\theta(t)\triangleq\left[\begin{array}{c}
\theta_{f}(t)\\
d(t)
\end{array}\right],\label{eq:theta}
\end{equation}

\noindent and the augmented regressor $Y:\mathbb{R}^{n}\times[0,\infty)\to\mathbb{R}^{n\times(n+m)}$
can be designed as 

\begin{equation}
Y(x(t),t)\triangleq\left[\begin{array}{cc}
Y_{h}(x(t),t) & I_{n}\end{array}\right].\label{eq:totparam}
\end{equation}

\noindent The parameterization in $\text{(\ref{eq:theta})}$ and $(\text{\ref{eq:totparam}})$
yields $h(x(t),t)+d(t)=Y(x(t),t)\theta(t)$, so the dynamics in $(\text{\ref{eq:dynamicsystem}})$
can be rewritten as 

\begin{equation}
\dot{x}(t)=Y(x(t),t)\theta(t)+u(t).\label{eq:xdotsub1}
\end{equation}

\begin{assumption}
\noindent The time-varying augmented parameter $\theta(t)$ and its
time-derivatives, i.e., $\dot{\theta}(t)$, $\ddot{\theta}(t)$ are
bounded by known constants, i.e., $\left\Vert \theta(t)\right\Vert \leq\bar{\theta}$,
$\left\Vert \dot{\theta}(t)\right\Vert \leq\zeta_{1},$ and $\left\Vert \ddot{\theta}(t)\right\Vert \leq\zeta_{2}$,
where $\bar{\theta},\zeta_{1},\zeta_{2}\in\mathbb{R}_{>0}$ are known
bounding constants, and $\left\Vert \cdot\right\Vert $ denotes the
Euclidean norm.
\end{assumption}

\section{Control Design}

\subsection{Control Objective}

The objective is to design a controller such that the state tracks
a smooth bounded reference trajectory, despite the time-varying nature
of the uncertain parameters. The objective is quantified by defining
the tracking error $e:[0,\infty)\to\mathbb{R}^{n}$ as\footnote{All function dependencies are suppressed equation $\text{\ensuremath{\left(\ref{eq:e}\right)}}$
onward; assume all variables to be time dependent unless stated otherwise.}

\begin{equation}
e\triangleq x-x_{d},\label{eq:e}
\end{equation}

\noindent where $x_{d}:[0,\infty)\to\mathbb{R}^{n}$ is a reference
trajectory.
\begin{assumption}
\noindent The reference trajectory $x_{d}(t)$ is bounded and smooth,
such that $\left\Vert x_{d}(t)\right\Vert \leq\bar{x}_{d}$, $\left\Vert \dot{x}_{d}(t)\right\Vert \leq\delta_{1}$,
and $\left\Vert \ddot{x}_{d}(t)\right\Vert \leq\delta_{2}$, where
$\bar{x}_{d},\delta_{1},\delta_{2}$$\in\mathbb{R}_{>0}$ are known
bounding constants.
\end{assumption}
Substituting $(\text{\ref{eq:xdotsub1}})$ into the time-derivative
of $(\text{\ref{eq:e}})$ yields
\begin{equation}
\dot{e}=Y\theta+u-\dot{x}_{d}.\label{eq:edotsub1}
\end{equation}

\noindent To facilitate the subsequent analysis, a filtered tracking
error $r:[0,\infty)\to\mathbb{R}^{n}$ is defined as 

\begin{equation}
r\triangleq\dot{e}+\alpha e,\label{eq:rdef}
\end{equation}

\noindent where $\alpha\in\mathbb{R}_{>0}$ is a constant control
gain. Substituting $(\text{\ref{eq:edotsub1}})$ into $(\text{\ref{eq:rdef}})$
yields

\begin{equation}
r=Y\theta+u-\dot{x}_{d}+\alpha e.\label{eq:r}
\end{equation}

\subsection{Control and Update Law Development}

From the subsequent stability analysis, the continuous control input
is designed as

\begin{equation}
u\triangleq-Y_{d}\hat{\theta}-\alpha e+\dot{x}_{d}+\mu,\label{eq:controlinput}
\end{equation}

\noindent where $Y_{d}\triangleq Y(x_{d}(t),t)$ is the desired regression
matrix, $\mu:[0,\infty)\to\mathbb{R}^{n}$ is a subsequently defined
auxiliary control term, and $\hat{\theta}:[0,\infty)\to\mathbb{R}^{n+m}$
denotes the parameter estimate of $\theta(t)$. Substituting the control
input in $(\text{\ref{eq:controlinput}})$ into the open-loop error
system in $(\text{\ref{eq:r}})$ yields the following closed-loop
system

\begin{equation}
r=Y\theta-Y_{d}\hat{\theta}+\mu.\label{eq:rsub1}
\end{equation}

\noindent Adding and subtracting $Y_{d}\theta$ in ($\text{\ref{eq:rsub1}}$)
yields

\begin{equation}
r=(Y-Y_{d})\theta+Y_{d}\tilde{\theta}+\mu,\label{eq:rsub2}
\end{equation}

\noindent where $\tilde{\theta}:[0,\infty)\to\mathbb{R}^{n+m}$ denotes
the parameter estimation error, i.e., $\tilde{\theta}(t)\triangleq\theta(t)-\hat{\theta}(t)$.
Taking the time-derivative of $(\text{\ref{eq:rsub2}})$ yields
\begin{alignat}{1}
\dot{r} & =(\dot{Y}-\dot{Y}_{d})\theta+(Y-Y_{d})\dot{\theta}+\dot{Y}_{d}\tilde{\theta}+Y_{d}\dot{\theta}-Y_{d}\dot{\hat{\theta}}+\dot{\mu}.\label{eq:rdot}
\end{alignat}

\noindent The control variables $\dot{\hat{\theta}}(t)$ and $\dot{\mu}(t)$
now appear in the higher order dynamics in $(\text{\ref{eq:rdot}}),$
and these control variables are designed with the use of a continuous
projection algorithm \cite[Appendix E]{Krstic1995}. The projection
algorithm constrains $\hat{\theta}(t)$ to lie inside a bounded convex
set $\mathcal{B}=\{\theta\in\mathbb{R}^{(n+m)}|\left\Vert \theta\right\Vert \leq\bar{\theta}\}$
by switching the adaptation law to its component tangential to the
boundary of the set $\mathcal{B}$ when $\hat{\theta}(t)$ reaches
the boundary. A continuously differentiable convex function $f:\mathbb{R}^{(n+m)}\to\mathbb{R}$
is used to describe the boundaries of the bounded convex set $\mathcal{B}$
such that $f(\theta(t))<0\,\,\mathbf{\forall}\,\left\Vert \theta(t)\right\Vert <\bar{\theta}$
and $f(\theta(t))=0\,\,\forall\,\left\Vert \theta(t)\right\Vert =\bar{\theta}$.
The adaptation law is then designed as

\begin{alignat}{1}
\dot{\hat{\theta}} & \triangleq\textrm{proj}(\Lambda_{0}(t))\nonumber \\
 & =\begin{cases}
\Lambda_{0}, & ||\hat{\theta}||<\bar{\theta}\,\lor\,(\nabla f(\hat{\theta}))^{T}\Lambda_{0}\leq0\\
\Lambda_{1}, & ||\hat{\theta}||\geq\bar{\theta}\,\land\,(\nabla f(\hat{\theta}))^{T}\Lambda_{0}>0,
\end{cases}\label{eq:adaptlaw}
\end{alignat}

\noindent where $||\hat{\theta}(0)||<\bar{\theta},$ $\lor$, $\land$
denote the logical `or', `and' operators, respectively, $\nabla$
represents the gradient operator, i.e., $\nabla f(\hat{\theta})$=$\left[\begin{array}{ccc}
\frac{\partial f}{\partial\phi_{1}} & \ldots & \frac{\partial f}{\partial\phi_{n+m}}\end{array}\right]_{\phi=\hat{\theta}}^{T}$, and $\Lambda_{0}:[0,\infty)\to\mathbb{R}^{n+m}$ and $\Lambda_{1}:[0,\infty)\to\mathbb{R}^{n+m}$
are designed as\footnote{\noindent From Lemma 1 in the Appendix section, $Y_{d}\Gamma Y_{d}^{T}$
is proven to be invertible, therefore it is reasonable to include
$(Y_{d}\Gamma Y_{d}^{T})^{-1}$ in the update law.}

\begin{equation}
\Lambda_{0}\triangleq\Gamma Y_{d}^{T}(Y_{d}\Gamma Y_{d}^{T})^{-1}\left[\beta\textrm{sgn}(e)\right],\label{eq:lambda0}
\end{equation}

\begin{equation}
\Lambda_{1}\triangleq\left(I_{m+n}-\frac{(\nabla f(\hat{\theta}))(\nabla f(\hat{\theta}))^{T}}{||\nabla f(\hat{\theta})||^{2}}\right)\Lambda_{0},\label{eq:lambda1}
\end{equation}

\noindent respectively. In $(\text{\ref{eq:lambda0}})$ and $(\text{\ref{eq:lambda1}})$,
$\beta\in\mathbb{R}_{>0}$ is a constant gain, and $\Gamma\in\mathbb{R}^{(n+m)\times(n+m)}$
is a positive-definite matrix with a block diagonal structure, i.e.,
$\Gamma\triangleq\left[\begin{array}{cc}
\Gamma_{1} & 0_{m\times n}\\
0_{n\times m} & \Gamma_{2}
\end{array}\right]$, with $\Gamma_{1}\in\mathbb{R}^{m\times m}$, $\Gamma_{2}\in\mathbb{R}^{n\times n}$.
The continuous auxiliary term $\mu(t)$, used in the control input
in $(\text{\ref{eq:controlinput}})$, acts as a stabilizing term in
the Lyapunov analysis to account for the side effects of the projection,
and is designed as a generalized solution to 
\begin{equation}
\dot{\mu}\triangleq\begin{cases}
\mu_{0}, & ||\hat{\theta}||<\bar{\theta}\,\lor\,(\nabla f(\hat{\theta}))^{T}\Lambda_{0}\leq0,\\
\mu_{1} & ||\hat{\theta}||\geq\bar{\theta}\,\land\,(\nabla f(\hat{\theta}))^{T}\Lambda_{0}>0,
\end{cases}\label{eq:riseterm}
\end{equation}

\noindent where $\mu(0)=0,$ and $\mu_{0}:[0,\infty)\to\mathbb{R}^{n}$
and $\mu_{1}:[0,\infty)\to\mathbb{R}^{n}$ are defined as $\mu_{0}\triangleq-Kr$
and $\mu_{1}\triangleq\mu_{0}-Y_{d}\left(\Lambda_{0}-\Lambda_{1}\right),$
respectively. Substituting $\left(\ref{eq:adaptlaw}\right)$ and $\left(\ref{eq:riseterm}\right)$
in $\left(\ref{eq:rdot}\right),$ the closed-loop dynamics can be
rewritten as

\begin{equation}
\dot{r}=(\dot{Y}-\dot{Y}_{d})\theta+(Y-Y_{d})\dot{\theta}+\dot{Y}_{d}\tilde{\theta}+Y_{d}\dot{\theta}-\beta\,\textrm{sgn}(e)-Kr,\label{eq:rdotsub2}
\end{equation}

\noindent for both cases, i.e., when $||\hat{\theta}||<\bar{\theta}\,\lor\,(\nabla f(\hat{\theta}))^{T}\Lambda_{0}\leq0$
or $||\hat{\theta}||\geq\bar{\theta}\,\land\,(\nabla f(\hat{\theta}))^{T}\Lambda_{0}>0.$
To facilitate the subsequent analysis, $(\text{\ref{eq:rdotsub2}})$
can be rewritten as
\begin{equation}
\dot{r}=\widetilde{N}+N_{B}-\beta\,\textrm{sgn}(e)-Kr-e,\label{eq:rdotsub3}
\end{equation}

\noindent where the variables $\widetilde{N}:[0,\infty)\to\mathbb{R}^{n}$
and $N_{B}:[0,\infty)\to\mathbb{R}^{n}$ are defined as 

\[
\widetilde{N}\triangleq(\dot{Y}-\dot{Y}_{d})\theta+(Y-Y_{d})\dot{\theta}+e,
\]

\noindent and 

\[
N_{B}\triangleq Y_{d}\dot{\theta}+\dot{Y}_{d}\theta-\dot{Y}_{d}\hat{\theta},
\]

\noindent respectively. The Mean Value Theorem (MVT) can be used to
develop the following upper bound on the term $\widetilde{N}(t)$

\begin{equation}
||\widetilde{N}||\leq\rho(||z||)||z||,\label{eq:ntildebound}
\end{equation}

\noindent where $z\triangleq\left[\begin{array}{cc}
r^{T} & e^{T}\end{array}\right]^{T}\in\mathbb{R}^{2n}$ and $\rho:\mathbb{R}^{2n}\to\mathbb{R}$ is a positive, globally
invertible and non-decreasing function. By Assumption 1, Assumption
2, Corollary 1 in the Appendix, and the bounding effect of projection
algorithm on $\hat{\theta}(t)$, the term $N_{B}(t)$ and its time-derivative
$\dot{N}_{B}(t)$ can be upper bounded by some constants $\gamma_{1},$$\gamma_{2}\in\mathbb{R}_{>0}$
as 

\begin{equation}
||N_{B}(t)||\leq\gamma_{1},\,\,\,\,\,\,\,\,\,\,\,\,\,\,\,\,\,\,\,\,||\dot{N}_{B}(t)||\leq\gamma_{2},\label{eq:nbbound}
\end{equation}
respectively.

\section{Stability Analysis}
\begin{thm}
The controller designed in $(\ref{eq:controlinput})$ along with the
adaptation laws designed in $\text{(\ref{eq:adaptlaw})}$ and $(\text{\ref{eq:riseterm}})$
ensure the closed-loop system is bounded and the tracking error $\left\Vert e(t)\right\Vert \to0$
as $t\to\infty,$ provided that the gains $\alpha,\beta$ are selected
such that the following condition is satisfied

\begin{equation}
\beta>\gamma_{1}+\frac{\gamma_{2}}{\alpha}.\label{eq:gaincondition}
\end{equation}
\end{thm}
\textit{Proof:} Let $\mathcal{D}\subset\mathbb{R}^{2n+1}$ be an open
connected set containing $y(t)=0$, where $y:[0,\infty)\to\mathbb{R}^{2n+1}$
is defined as 

\[
y(t)\triangleq\left[\begin{array}{cc}
z^{T}(t) & \sqrt{P(t)}\end{array}\right]^{T}.
\]

\noindent Let $V_{L}:\mathcal{D}\times[0,\infty)\to\mathbb{R}_{\geq0}$
be a positive-definite candidate Lyapunov function defined as

\[
V_{L}(y(t),t)\triangleq\frac{1}{2}r^{T}r+\frac{1}{2}e^{T}e+P,
\]

\noindent where $P:[0,\infty)\to\mathbb{R}$ is a generalized solution
to the differential equation

\begin{equation}
\text{\ensuremath{\dot{P}}}(t)\triangleq-L(t),\,\label{eq:Pdot}
\end{equation}

\noindent where $P(0)\triangleq\mathop{\beta\stackrel[i=1]{n}{\sum}\left|e_{i}(0)\right|-e(0)^{T}N_{B}(0)}$
and 
\begin{equation}
L\triangleq r^{T}(N_{B}-\beta\textrm{sgn}(e)).\label{eq:L}
\end{equation}

\begin{rem}
Provided that the gain condition in $(\text{\ref{eq:gaincondition}})$
is satisfied, $P(t)\geq0$.\footnote{See \cite{Xian2004} for details.}
Hence it is valid to use $P(t)$ in the candidate Lyapunov function
as function of the variable $\sqrt{P(t)}$. 
\end{rem}
$ $

\noindent From $(\text{\ref{eq:rdef}})$, ($\text{\ref{eq:rdotsub3}}$)
and $(\text{\ref{eq:Pdot}})$, the differential equations describing
the closed-loop system are \vspace{-3pt}
\begin{eqnarray}
\dot{e} & = & r-\alpha e,\label{eq:edotsub2}\\
\dot{r} & = & \widetilde{N}+N_{B}-\beta\,\textrm{sgn}(e)-Kr-e,\label{eq:rdotsub4}\\
\dot{P} & = & -r^{T}(N_{B}-\beta\textrm{sgn}(e)).\label{eq:Pdotsub1}
\end{eqnarray}

\noindent Let $g:\mathbb{R}^{2n+1}\times[0,\infty)\to\mathbb{R}^{2n+1}$
denote the right-hand side of $(\text{\ref{eq:edotsub2}})$-$(\text{\ref{eq:Pdotsub1}})$.
Since $g(y(t),t)$ is continuous almost everywhere, except in the
set $\{(y(t),t)|e=0\}$, an absolute continuous Filippov solution
$y(t)$ exists almost everywhere (a.e.), so that $\dot{y}(t)\in K[g](y(t),t)$
a.e., except at the points in the set $\{(y(t),t)|e=0\}$, where the
Filippov set-valued map includes unique solutions. Using a generalized
Lyapunov stability theory under the framework of Filippov solutions,
a generalized time-derivative of the Lyapunov function $V_{L}$ exists
and $\dot{V}_{L}(y,t)\in\dot{\widetilde{V}}_{L}(y,t)$ , where

\begin{eqnarray}
\dot{\widetilde{V}}_{L}(y,t) & = & \underset{\xi\in\partial V_{L}(y,t)}{\bigcap}\xi^{T}K\left[\begin{array}{ccc}
\dot{e}^{T} & \dot{r}^{T} & \frac{1}{2}P^{-\frac{1}{2}}\dot{P}\end{array}\right]^{T}\nonumber \\
 & = & \nabla V_{L}^{T}K\left[\begin{array}{ccc}
\dot{e}^{T} & \dot{r}^{T} & \frac{1}{2}P^{-\frac{1}{2}}\dot{P}\end{array}\right]^{T}\nonumber \\
 & \subset & \left[\begin{array}{ccc}
e^{T} & r^{T} & 2P^{\frac{1}{2}}\end{array}\right]\times\nonumber \\
 &  & K\left[\begin{array}{ccc}
\dot{e}^{T} & \dot{r}^{T} & \frac{1}{2}P^{-\frac{1}{2}}\dot{P}\end{array}\right]^{T},\label{eq:vltildedotsub2}
\end{eqnarray}

\noindent where $\partial V_{L}(y,t)$ denotes Clarke's generalized
gradient \cite{Fischer.Kamalapurkar.ea2013}. Substituting $(\text{\ref{eq:edotsub2}})$-$(\text{\ref{eq:Pdotsub1}})$
into $(\text{\ref{eq:vltildedotsub2}})$ yields

\begin{alignat}{1}
\dot{\widetilde{V}}_{L} & \overset{a.e.}{\subset}r^{T}(\widetilde{N}+N_{B}-\beta\,\textrm{sgn}(e)-Kr-e)\nonumber \\
 & +e^{T}(r-\alpha e)-r^{T}(N_{B}-\beta\textrm{sgn}(e))\label{eq:vltildedotsub1}
\end{alignat}

\noindent where $K\left[\textrm{sgn}(e)\right]=\textrm{SGN}(e)$ such
that

\[
\textrm{SGN}(e_{i})=\begin{cases}
\left\{ 1\right\} , & e_{i}>0\\
\left[-1,1\right], & e_{i}=0\\
\left\{ -1\right\} , & e_{i}<0.
\end{cases}
\]

\noindent Using $\text{\ensuremath{\left(\ref{eq:ntildebound}\right)}}$,
the expression in $\left(\text{\ref{eq:vltildedotsub1}}\right)$ can
be upper bounded as

\[
\dot{\widetilde{V}}_{L}\overset{a.e.}{\leq}\rho(\left\Vert z\right\Vert )\left\Vert z\right\Vert \left\Vert r\right\Vert -K\left\Vert r\right\Vert ^{2}-\alpha e^{2}.
\]
Using Young's Inequality on $\rho(\left\Vert z\right\Vert )\left\Vert z\right\Vert \left\Vert r\right\Vert $
yields $\rho(\left\Vert z\right\Vert )\left\Vert z\right\Vert \left\Vert r\right\Vert \leq\frac{\rho^{2}(\left\Vert z\right\Vert )\left\Vert z\right\Vert ^{2}}{2}+\frac{1}{2}\left\Vert r\right\Vert ^{2}$.
Therefore,

\begin{eqnarray}
\dot{\widetilde{V}}_{L} & \overset{a.e.}{\leq} & \frac{\rho^{2}(\left\Vert z\right\Vert )\left\Vert z\right\Vert ^{2}}{2}-(K-\frac{1}{2})\left\Vert r\right\Vert ^{2}-\alpha e^{2}\nonumber \\
 & \overset{a.e.}{\leq} & -\left(\lambda_{3}-\frac{\rho^{2}(\left\Vert z\right\Vert )}{2}\right)\left\Vert z\right\Vert ^{2},\label{eq:VLtdot}
\end{eqnarray}

\noindent where $\lambda_{3}\triangleq\min\{\alpha,K-\frac{1}{2}\}\in\mathbb{R}_{>0}$
is a known constant. The expression in $(\text{\ref{eq:VLtdot})}$
can be rewritten as 

\begin{equation}
\dot{V}_{L}\overset{a.e.}{\leq}-c\left\Vert z\right\Vert ^{2}\,\forall\,y\in\mathcal{D},\label{eq:VLdot}
\end{equation}

\noindent for some constant $c\in\mathbb{R}_{>0}$, where 

\[
\mathcal{D}\triangleq\left\{ y\in\mathbb{R}^{2n+1}|\left\Vert y\right\Vert \leq\rho^{-1}\left(\sqrt{2\lambda_{3}}\right)\right\} .
\]

In this region, $\lambda_{3}>\frac{\rho^{2}(\left\Vert z\right\Vert )}{2}$,
so a constant $c$ satisfies $\text{(\ref{eq:VLdot})}$, and larger
values of $\lambda_{3}$ expand the size of $\mathcal{D}.$ Furthermore,
the relationship in $(\ref{eq:VLdot})$ implies that $V_{L}(y(t),t)\in\mathcal{L}_{\infty},$
hence $e(t),r(t)$, $P(t)$ $\in\mathcal{L}_{\infty}$. These facts
along with the expression in $(\text{\ref{eq:rsub2}})$, indicate
that $\mu(t)\in\mathcal{L}_{\infty}$. The parameter estimate $\hat{\theta}(t)\in\mathcal{L}_{\infty}$
due to the projection operation. The state and its time-derivative,
i.e., $x(t),\dot{x}(t)\in\mathcal{L}_{\infty}$, because $e(t),r(t),x_{d}(t),\dot{x}_{d}(t)\in\mathcal{L}_{\infty}.$
Further the regression matrix $Y(x(t),t)\in\mathcal{L}_{\infty}$
since its a bounded function for a bounded argument $x(t)$. Similarly,
$Y_{d}(t)\in\mathcal{L}_{\infty}$, hence $\dot{\hat{\theta}}\in\mathcal{L}_{\infty}$
by Corollary 1. From the expression in $\text{(\ref{eq:controlinput}}$),
since $\hat{\theta}(t),e(t),\dot{x}_{d}(t),\mu(t)\in\mathcal{L}_{\infty}$,
$u(t)\in\mathcal{L}_{\infty}$. Hence all the closed-loop signals
are bounded.

$ $

Consider $\lambda_{1}\left\Vert y\right\Vert ^{2}\leq V_{L}\leq\lambda_{2}\left\Vert y\right\Vert ^{2},$
where $\lambda_{1},\lambda_{2}\in\mathbb{R}_{>0}$. To ensure $\left\Vert z\right\Vert \leq\rho^{-1}(\sqrt{2\lambda_{3}})$,
it is sufficient to obtain the result from $\left\Vert y\right\Vert \leq\rho^{-1}(\sqrt{2\lambda_{3}})$.
Since $\sqrt{\frac{V_{L}}{\lambda_{2}}}\leq\left\Vert y\right\Vert $,
then $\sqrt{\frac{V_{L}}{\lambda_{2}}}\leq\rho^{-1}(\sqrt{2\lambda_{3}})$,
and $V_{L}$ is non-increasing, so $V_{L}(t)\leq V_{L}(0)$. Hence
it sufficient to show that $\sqrt{\frac{V_{L}(0)}{\lambda_{2}}}\leq\rho^{-1}(\sqrt{2\lambda_{3}})$
to ensure that $\sqrt{\frac{V_{L}}{\lambda_{2}}}\leq\rho^{-1}(\sqrt{2\lambda_{3}})$.
Since $\lambda_{1}\left\Vert y(0)\right\Vert ^{2}\leq V_{L}(0)$ implies
$\left\Vert y(0)\right\Vert \leq\sqrt{\frac{V_{L}(0)}{\lambda_{1}}}\leq\sqrt{\frac{\lambda_{2}}{\lambda_{1}}}\rho^{-1}(\sqrt{2\lambda_{3}}),$
so $y\in\mathcal{S}\triangleq\left\{ y(t)\in\mathcal{D}|y(t)\leq\sqrt{\frac{\lambda_{2}}{\lambda_{1}}}\rho^{-1}(\sqrt{2\lambda_{3}})\right\} $
is the region where $y(0)$ should lie for guaranteed asymptotic stability.
The gain condition $\lambda_{3}=\min\{\alpha,K-\frac{1}{2}\}\geq\frac{\rho^{2}\left(\sqrt{\frac{\lambda_{1}}{\lambda_{2}}}\left\Vert y(0)\right\Vert \right)}{2}$
needs to be satisfied according to the initial condition for asymptotic
stability and the region of attraction can be made arbitrarily large
to include any initial condition by increasing the gains $\alpha$
and $K$ accordingly. By the extension of LaSalle-Yoshizawa theorem
for non-smooth systems in \cite{Fischer.Kamalapurkar.ea2013} and
\cite{Kamalapurkar.Rosenfeld.ea2019}, $c\left\Vert z(t)\right\Vert ^{2}\to0$
and hence $\left\Vert e\right\Vert \to0$ as $t\to\infty\,\,$$\forall\,\,y(0)\in\mathcal{S}$
, so the closed-loop error system is semi-globally asymptotically
stable.
\begin{flushright}
$\blacksquare$
\par\end{flushright}

\section{Conclusion}

A continuous adaptive control design was presented to achieve semi-global
asymptotic tracking for linearly parameterizable nonlinear systems
with time-varying uncertain parameters. The key feature of this design
is a RISE-like parameter update law along with a projection algorithm,
which allows the system to compensate for potentially destabilizing
terms in the closed-loop error system, arising due to the time-varying
nature of parameters. Semi-global asymptotic tracking for the error
system is guaranteed via a Lyapunov-based stability analysis. Future
work will involve improvement of the parameter estimation performance
of time-varying parameter systems and its extension to the system
identification problem. 

\bibliographystyle{IEEEtran}
\bibliography{ncr,master,1C__Users_patilomkarsudhir_Desktop_Papers_TAC_2___of_Time_Varying_Parameter_Systems_CustomBib}

\appendix{}
\begin{lem}
\noindent Consider a positive-definite matrix $\Gamma\in\mathbb{R}^{(n+m)\times(n+m)}$
such that $\Gamma$ has the block diagonal structure as $\Gamma\triangleq\left[\begin{array}{cc}
\Gamma_{1} & 0_{m\times n}\\
0_{n\times m} & \Gamma_{2}
\end{array}\right]$, where $\Gamma_{1}\in\mathbb{R}^{m\times m}$ and $\Gamma_{2}\in\mathbb{R}^{n\times n}$.
The matrix $Y(x(t),t)\Gamma Y^{T}(x(t),t)$ is positive-definite,
and hence invertible. Furthermore, the inverse of this matrix satisfies
the property $\left\Vert \left(Y(x(t),t)\Gamma Y^{T}(x(t),t)\right)^{-1}\right\Vert _{2}\leq\frac{1}{\lambda_{\textrm{min}}\left\{ \Gamma_{2}\right\} }$,
where $\left\Vert \cdot\right\Vert _{2}$ denotes the spectral norm
and $\lambda_{\textrm{min}}\left\{ \cdot\right\} $ denotes the minimum
eigenvalue of $\left\{ \cdot\right\} $. 
\end{lem}
\textit{Proof : }Substituting the definitions for $Y(x(t),t)$ and
$\Gamma$ in $Y(x(t),t)\Gamma Y^{T}(x(t),t)$ yields
\begin{align*}
Y(x(t),t)\Gamma Y^{T}(x(t),t) & =\\
\left[\begin{array}{cc}
Y_{h}(x(t),t) & I_{n}\end{array}\right]\left[\begin{array}{cc}
\Gamma_{1} & 0_{m\times n}\\
0_{n\times m} & \Gamma_{2}
\end{array}\right] & \left[\begin{array}{c}
Y_{h}(x(t),t)\\
I_{n}
\end{array}\right]\\
=Y_{h}(x(t),t)\Gamma_{1}Y_{h}(x(t),t) & +\Gamma_{2}.
\end{align*}
Since $\Gamma$ is selected to be a positive-definite matrix, the
block matrices $\Gamma_{1}$ and $\Gamma_{2}$ are both positive-definite,
so the first term $Y_{h}(x(t),t)\Gamma_{1}Y_{h}(x(t),t)$ in this
expression is positive semi-definite while the second term $\Gamma_{2}$
is positive-definite, hence the sum of these two terms, i.e., $Y(x(t),t)\Gamma Y^{T}(x(t),t)$
is positive-definite, and therefore invertible. Furthermore, the spectral
norm satisfies the property, $\left\Vert A\right\Vert _{2}=\sqrt{\lambda_{\textrm{max}}\left\{ A^{T}A\right\} }$
for some $A\in\mathbb{R}^{p\times q}$ with $p,q\in\mathbb{Z}_{>0}$,
where $\lambda_{\textrm{max}}\left\{ \cdot\right\} $ denotes the
maximum eigenvalue of $\left\{ \cdot\right\} $. Utilizing this property
with $\left\Vert \left(Y\Gamma Y^{T}\right)^{-1}\right\Vert _{2}$
yields 

\noindent 
\begin{align}
\left\Vert \left(Y\Gamma Y^{T}\right)^{-1}\right\Vert _{2} & =\sqrt{\lambda_{\textrm{max}}\left\{ \left(\left(Y\Gamma Y^{T}\right)^{-1}\right)^{T}\left(Y\Gamma Y^{T}\right)^{-1}\right\} }\nonumber \\
 & =\lambda_{\textrm{max}}\left\{ \left(Y\Gamma Y^{T}\right)^{-1}\right\} .\label{eq:lemma1-1}
\end{align}

\noindent The eigenvalues of the inverse of a positive definite matrix
$B$ satisfy the property, $\lambda_{\textrm{max}}\left\{ B^{-1}\right\} =\frac{1}{\lambda_{\textrm{min}}\left\{ B\right\} }$.
Applying this property with the right-hand side of $(\text{\ref{eq:lemma1-1}})$
yields 

\begin{eqnarray*}
\left\Vert \left(Y\Gamma Y^{T}\right)^{-1}\right\Vert _{2} & = & \frac{1}{\lambda_{\textrm{min}}\left\{ Y\Gamma Y^{T}\right\} }\\
 & \leq & \frac{1}{\lambda_{\textrm{min}}\left\{ \Gamma_{2}\right\} }.
\end{eqnarray*}

\noindent \begin{flushright}
$\blacksquare$
\par\end{flushright}
\begin{cor}
The norm of time-derivative of the parameter estimate, $\left\Vert \dot{\hat{\theta}}\right\Vert $
can be upper bounded by a constant $\gamma_{3}\in\mathbb{R}_{>0}$,
i.e., $\left\Vert \dot{\hat{\theta}}\right\Vert \leq\gamma_{3}$.
\end{cor}
\begin{IEEEproof}
Based on $(\text{\ref{eq:adaptlaw}})$

\begin{eqnarray}
\left\Vert \dot{\hat{\theta}}\right\Vert  & = & \left\Vert \textrm{proj}(\Lambda_{0})\right\Vert \leq\left\Vert \Lambda_{0}\right\Vert \nonumber \\
 & = & \left\Vert \Gamma Y_{d}^{T}(Y_{d}\Gamma Y_{d}^{T})^{-1}\beta\textrm{sgn}(e)\right\Vert \nonumber \\
 & \leq & \left\Vert \Gamma Y_{d}^{T}(Y_{d}\Gamma Y_{d}^{T})^{-1}\beta\right\Vert .\label{eq:corr1}
\end{eqnarray}
Applying Holder's inequality to the right-hand side of $(\text{\ref{eq:corr1}})$
yields
\begin{equation}
\left\Vert \dot{\hat{\theta}}\right\Vert \leq\beta\left\Vert \Gamma\right\Vert _{2}\left\Vert Y_{d}\right\Vert _{2}\left\Vert (Y_{d}\Gamma Y_{d}^{T})^{-1}\right\Vert _{2}.\label{eq:corr2}
\end{equation}
Using Lemma 1 with the right-hand side of $(\text{\ref{eq:corr2}})$
yields

\[
\left\Vert \dot{\hat{\theta}}\right\Vert \leq\frac{\beta\left\Vert \Gamma\right\Vert _{2}\left\Vert Y_{d}\right\Vert _{2}}{\lambda_{\textrm{min}}\left\{ \Gamma_{2}\right\} }.
\]
Given a bounded reference $x_{d}(t)$, such that $\left\Vert x_{d}(t)\right\Vert \leq\bar{x}_{d}$,
the spectral norm of the desired regressor may be upper-bounded by
a constant $\bar{Y}_{d}\in\mathbb{R}_{>0}$, i.e., $\left\Vert Y_{d}\right\Vert _{2}\leq\bar{Y}_{d},$
because $Y_{d}$ is a continuously differentiable function. Therefore,
\[
\left\Vert \dot{\hat{\theta}}\right\Vert \leq\frac{\beta\left\Vert \Gamma\right\Vert _{2}\bar{Y}_{d}}{\lambda_{\textrm{min}}\left\{ \Gamma_{2}\right\} }=\gamma_{3}.
\]
\end{IEEEproof}

\end{document}